# ELECTRONIC EMITION PROPERTIES OF BI LAYER NOVEL ORGANIC SEMICONDUCTOR SYSTEMS

*Salazar Valencia Pablo.  **Bolívar Marinez Luz Elena.  ***Pérez Merchancano Servio
* pjsava@gmail.com, ** lbolivar@unicauca.edu.co, *** sperez@unicauca.edu.co
Departamento de Física.  Universidad del Cauca. Calle 5 # 4-70.  Campus Tulcán.  Popayán, Colombia

The perylene-3,4,9,10-tetracarboxylic-dianhydride (PTCDA) and 1,4,5,8-naphthalene-tetracaboxylic-dianhydride (NTCDA) are planar π-stacking organic molecules that have been shown to be excellent model compounds for studying the growth and optoelectronic properties of organic semiconductor thin films, particularly organic diodes.  Some observations have shown that this molecules, particularly PTCDA a brick-like shaped molecule easily forms well-ordered films on various substrates due to its unique crystal structure which is characterized by flat lying molecules In this work we will explore some energetic and optical characteristics such as heats of formation, optic GAP energies, electronic transitions and others of novel tow layer systems of alternate layers of PTCDA and NTCDA by means of the semiempirical methods Parametric Model 3 (PM3) and Zerner's Intermediate Neglect of Differential Overlap (ZINDO/S) in Configuration Interaction mode.

1.  Introduction and Methodology

PTCDA and NTCDA (Fig 1) are planar π-stacking organic molecules widely used for the study of the epitaxial growth of organic semiconductor thin films. Both molecules share common structural characteristics consisting of a perylene core the PTCDA and a naphthalene the NTCDA with a delocalized π electron system and two anhydride terminal groups. This gives rise to a permanent quadrupole moment with the positive charge situated around the center of the molecule, and the negative charge around the terminal groups, its dimensions are respectively of 14 Å in length and 9.2 Å width for PTCDA and 7.12 Å in length and 5.1 Å in width for the NTCDA both with monoclinic crystalline structures [1].

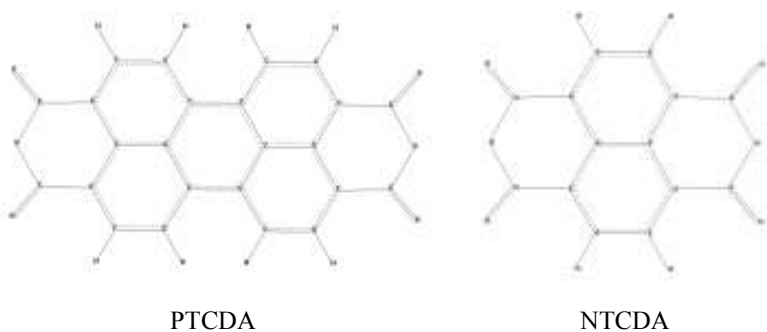

PTCDA            NTCDA

Fig. 1. Structures of PTCDA and NTCDA.

All the structures studied in this work have been previously obtained by a geometric optimization using the semiempirical method, Parametric Model 3 of the NDO family. Given the poor quality in the treatment of the excited states and electron correlation which is inherent to the NDO family methods, the simulation of the electronic or ultraviolet spectra will be obtained with the Zerner's spectroscopic version of the Intermediate

Neglect of Differential Overlap in configuration interactions mode: ZINDO/S CI, with a 3 occupied and 3 unoccupied orbital criterion. The calculations were performed with the computational package HyperChem 5.11 of the Hypercube house. All these methodologies have been extensible tested and are known to be pretty accurate and reliable in the study of organic molecules [2].

2. Results and Discussion

In this work we have studied 2 layer systems of PTCDA and NTCDA in comparison with the isolated molecules with the porpoise of compare its optical behavior under ideal conditions, according to the results of the simulations as seen in the Table 1, the heat of formation values from the PM3 Semiempirical Method cover a range from -1748.248 kcal/mol for the PTCDA to the -334.225 kcal/mol for the tow layer system of NTCDA and PTCDA being this one the lowest found [3].

Table 1. Heat of formation values in kcal/mol, HOMO, LUMO, and Optical GAP energies in electron volts, and most intense optical transitions for the structures studied in this work, all results from the geometry optimizations with the semiempirical methods PM3 and ZINDO/S CI.

|  | Heat of Formation | HOMO | LUMO | GAP | TRANSTIONS |
| --- | --- | --- | --- | --- | --- |
| NTCDA | -186,188 | -9.48 | -2.48 | 7.00 | H→L+2: 0.388 <br> H-1→L: 0.591 |
| PTCDA | -148.248 | -8.13 | -2.49 | 5.64 | H→L: 0.706 |
| 2 NTCDA | -332.939 | -7.40 | 0.74 | 8.14 | H→L+1: 0.707 |
| 2 PTCDA | -298.003 | -8.10 | -2.48 | 5.62 | H→L+1: 0.501 <br> H-1→L: 0.498 |
| NTCDA + PTCDA | -334.225 | -8.18 | -2.59 | 5.59 | H→L: 0.707 |

The simulation of the ultraviolet spectra obtained with the ZINDO/S CI semiempirical method are presented in the Figure 2, with some complementary information also on the Table 1, for the isolated systems the conduction begins in the range from 3.1 to 4.4 eV and the most intense peaks are located for the PTCDA in the 3 eV region, and for the NTCDA in the 6eV region.

For the combined systems the behavior is more variable with the conduction beginning earlier in the region of the 2 eV with the 2 layer NTCDA in the 3 eV region, the 2 layer PTCDA in the 6 eV region and the NTCDA – PTCDA combination beginning almost in the limit of the 2eV showing a much better conducting behavior that of the isolated structures; this results are pretty close from the experimental ones that locate the isolated PTCDA and NTCDA structures in the 2.2 eV region [4, 5].

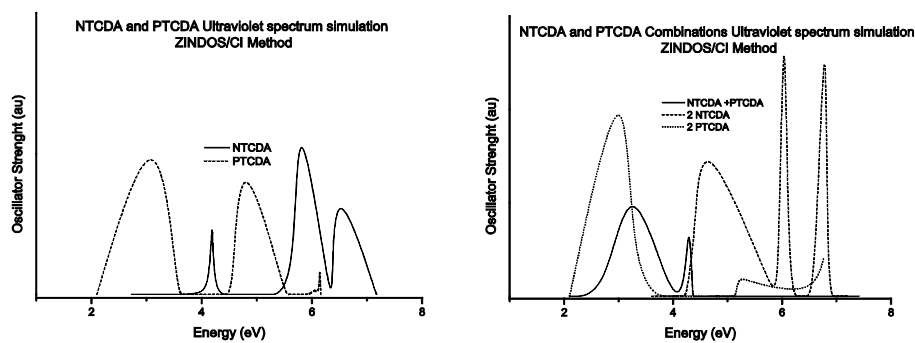

Fig. 2. Ultraviolet spectrum simulation of the structures studied in this work, simulations obtained form geometry optimizations with the semiempirical methods PM3 and ZINDO/S CI, energy in electron volts and oscillator strength in arbitrary units.

The most intense optical transitions for all the structures involve only the frontier orbitals spanning the range from one orbital below the HOMO and two orbitals above the LUMO, also from the data in table 1 we can see that the optical GAP values for the two layer systems tend to diminish being the NTCDA and PTCDA system the one with the lowest value, this behavior can show a tendency to decrease the optical GAP as more alternate layers of the tow molecular systems studied here are added.  A further observation of the frontier orbitals has shown an interesting phenomenon in the two layer system of PTCDA and NTCDA, in this system the HOMO orbital is restricted only to the PTCDA layer, while the NTCDA layer contains the LUMO orbital showing that the conduction processes occur now from one system to the other.  For the all the other molecular structures studied in this paper the transition HOMO → LUMO involves all the components in the system [6].

4. Conclusions

The most interesting phenomena observed in this work is the change in the optical GAP values for the two layers systems of PTCDA and NTCDA form those observed in the isolated structures, this system exhibits the lowest heat of formation values and the lowest optical GAP indicating a tendency to improve the conducting properties of such systems, a further study is necessary to confirm if the properties here observed are maintained in bigger systems.

3.  References

[1]. S.R. Forrest, M.L.Kaplan, y P.H. Schmidt, J.Appl. Phys. (1984); 149, 255 (6), J.Appl. Phys. (1984), 543, 56 (2).

[2] H. Fuchigami, S. Tanimura, y S. Tsunoda, Jpn. J. Appl. Phys. (19950, 3852, 34,

[3] P.E. Burrows y S.R. Forrest, Appl. Phys. Lett. (1994) 2285. 64 (17), A. Ardavan, S.J.Blundell, J. Singleton, ArXiv: cond-mat/0110280v1, 2001.

[4] J.J.P. Stewart, J. comp. Chem.. (1989)  10,. Mopac and Spartan: Hyperchem 5.1 SN500-10005433.

[5]J.Ridley, M.Zerner, Theor.Chem. Acta, (1987), 347, 72.

[6] D.S.Galvao, D.A.dos Santos, B.Laks, M.J. Caldas, Phys. Rev. Lett. (1989), 63, 786.